\documentclass[12pt,a4paper,reqno]{amsart}
\usepackage{graphicx} 
\graphicspath{{figures/}} 
\usepackage{amsmath,amssymb,amsthm,array}
\usepackage[font=small,labelfont=bf]{caption} 
\usepackage{amsfonts, amsmath, amsthm, amssymb} 
\usepackage{wrapfig} 

\begin{document}

\title{On nonlinear superposition \\ of the KdV-Burgers equation}

\author{Alexey Samokhin}\vspace{6pt}

\address{Dept. of Math., Moscow State Technical University of Civil Aviation}\vspace{6pt}

\email{ samohinalexey@gmail.com}\vspace{6pt}


\begin{abstract} Superposition of explicit (analytic) monotone non-increasing shock waves for the KdV-Burgers equation is studied and modelled numerically.
Initial profile chosen as a sum of two such shock waves gradually transforms into a single shock wave of a somewhat complex yet predictable structure. This transformation is demonstrated in detail. \vspace{1mm}

\noindent\textbf{Keywords:} nonlinear superposition,invariant solution, KdV-Burgers equation.
\end{abstract}

\maketitle

\normalsize\section{Introduction}

The solutions to the KdV-Burgers equation are an object of intensive study for years. However the their asymptotic at $t\rightarrow\infty$ were studied on the whole line ,  \cite{key-6,key-4}. Lately the studies are focused on such aspects as gradient catastrophes, breaks, critical points, etc, \cite{key-1,key-2,key-3,key-5}.

In this paper a superposition of explicit (analytic) monotone non-increasing shock waves for the KdV-Burgers equation (kdV-B)
\begin{equation}\label{eq}
u_t=\varepsilon^2 u_{xx}-2uu_x+\lambda u_{xxx}
\end{equation}
 is studied and modelled numerically on a finite interval $x\in [0,\, L]$ for suitably large  $L$; $\lambda$ and $\varepsilon$ are the dispersion and dissipation coefficients respectively.

 The initial value - boundary problem is of the form
 \begin{equation}\label{IVBP}
u(x,0) =f(x,0), \quad u(a,t) = H, \quad u_x(a,t) = u_x(b,t) =0, \quad x\in[a,b],
\end{equation}
where $f(x,t)=TWS_1(x-V_1t)+TWS_2(x-V_2t)$
is a sum of two monotone non-increasing shock waves (TWS stands for a "Traveling Wave Solution").

\section{Invariant solutions}

We  recall some facts.
Some of the shock TWS's for KdV-B have the following explicit form:

\begin{equation}\label{TWS}
TWS=\frac{3\varepsilon^4\tanh^2(\frac{\varepsilon^2(x-Vt-s)}{10\lambda})}{50\lambda} -
\frac{3\varepsilon^4\tanh(\frac{\varepsilon^2(x-Vt-s)}{10\lambda})}{25\lambda}+\frac{V}{2}-\frac{3\varepsilon^4}{50\lambda}
\end{equation}

\begin{figure}[h]
\begin{minipage}{13.2pc}
\includegraphics[width=13.2pc]{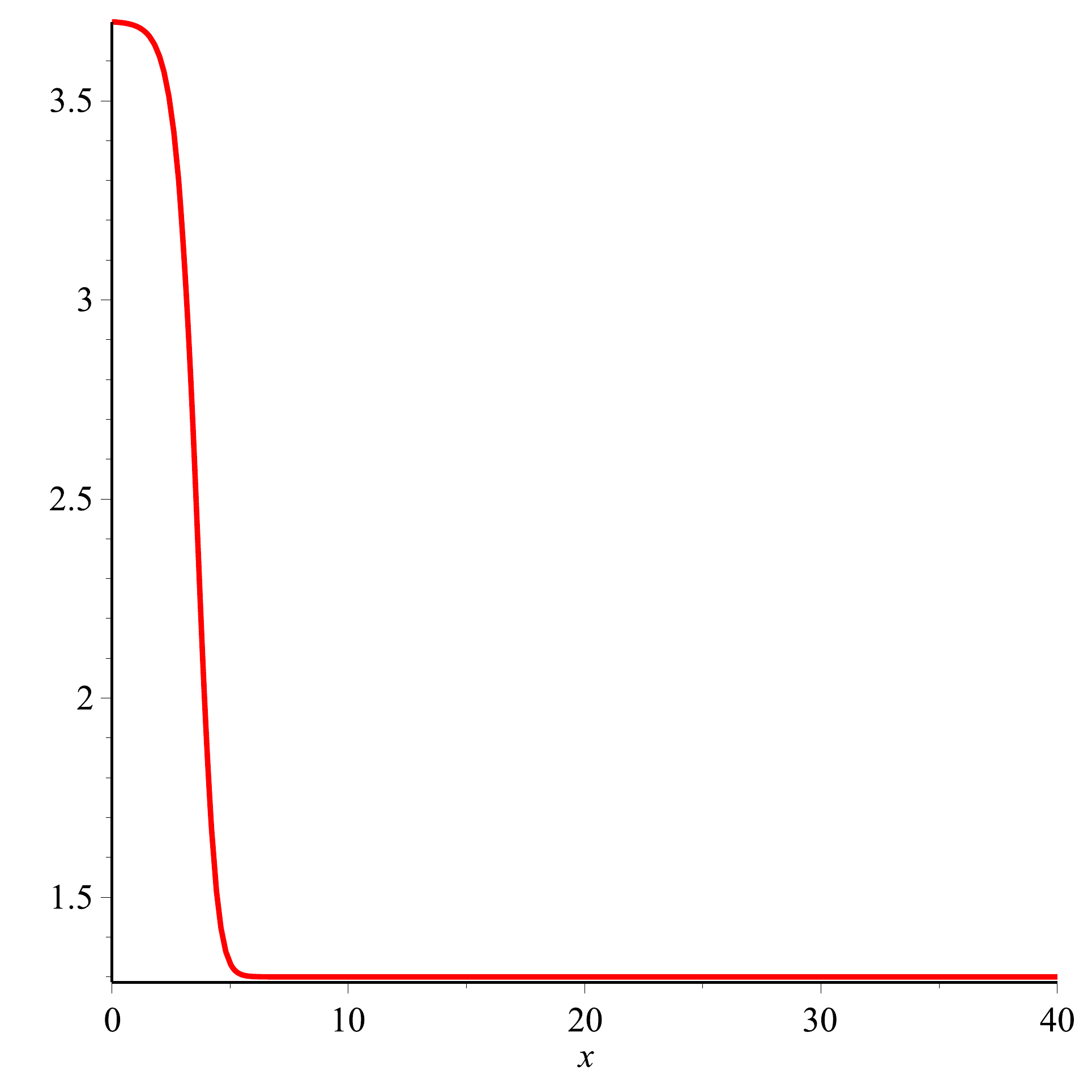}
\end{minipage}\hspace{2pc}%
\begin{minipage}{13.2pc}
\includegraphics[width=13.2pc]{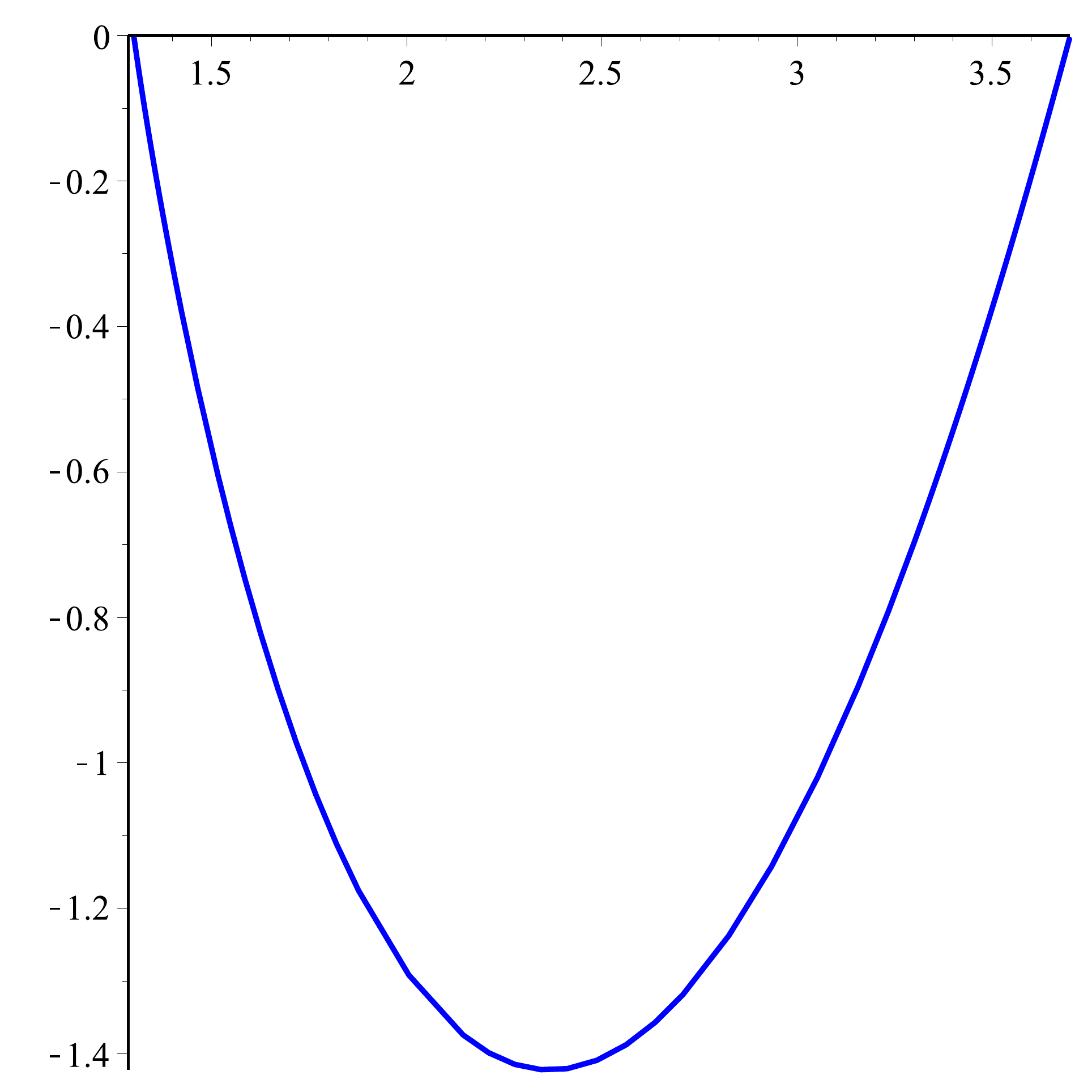}
\end{minipage}
\caption{\label{g1}Invariant shock wave and its phase portrait.\\
$\lambda=0.1,\;\varepsilon=1,\;V= 5,\; s=4,\;t=0$}
\end{figure}

 Here $s$ is a shift, $V$ --- a velocity of the wave; the typical profile of such a wave is resented on figure \ref{g1}, left. On the same figure the wave phase portrait is also presented on the right.

For analytical solutions of the form (\ref{TWS}) it is esy to find their limit levels $H$ and $h$:
\[\left.TWS\right|_{\xi=\pm\infty}=\frac{V}{2}\pm \frac{3\varepsilon^4}{25\lambda}.\]
Thus,
 the velocity of the shock wave  \eqref{TWS} connected with  the limit values by the formula $V=H+h$.

 Note also that the height of the wave,
 $H-h=6\varepsilon^4/25\lambda$, does not depend on  the velocity and is entirely  defined by  the ratio of the $\varepsilon^4$ and $\lambda$ coefficients, linked to dissipation and dispersion.

All solution of the form $u(x,t)=y(x-Vt)=y(\xi)$ satisfy the ordinary differential equation

 \[
 \lambda y'''+\varepsilon^2 y''-2yy'+Vy'=0,\quad y'=\frac{dy}{d\xi},
 \]

 whose order may be lowered:
 \[\lambda y''+\varepsilon^2 y'-y^2-Vy+C=0.
 \]
Rewrite the latter as a dynamical system

 \begin{equation}\label{sys}
   \left\{\begin{array}{rcl}
     y' &=&p \\
     p' &=&\frac{1}{\lambda} (-\varepsilon^2 p+y^2-Vy+C)
   \end{array}\right.
 \end{equation}

Note that for the explicit solutions (\ref{TWS}) the integration constant $C$ is readily calculated  \[C=\frac{V^2}{4}-\frac{9\varepsilon^8}{625\lambda^2}.\]
 Hence the critical points $(y,p)$ of the system \eqref{sys} correspond to $p=0$ and  $y$ are the roots of the quadratic equation
 \[y^2-Vy+\frac{V^2}{4}-\frac{9\varepsilon^8}{625\lambda^2}=0.
 \]
These roots are
 \[U_\pm=\frac{V}{2}\pm \frac{3\varepsilon^4}{25\lambda};\]
 they (naturally) coincide with the limit values  $H,\,h$.

 To find the types of these critical points on the phase plane $(y,p)$ it is necessary  to find the eigenvalues for the linearisation of the system \eqref{sys}:

 \begin{equation}\label{lin}
   \left(
     \begin{array}{cc}
       0 & 1 \\
     \frac{2y-V}{\lambda}   & -\frac{\varepsilon^2}{\lambda} \\
     \end{array}
   \right)
 \end{equation}

 Characteristic equation is as follows

 \begin{equation}\label{qe}
 k^2+\frac{\varepsilon^2}{\lambda}k-\frac{2y-V}{\lambda}=0.
 \end{equation}

  At the point $p=0$,
 $y=\frac{V}{2}- \frac{3\varepsilon^4}{25\lambda}$ the roots of \eqref{qe} are  $k_\pm=\frac{\varepsilon^2}{2\lambda}\left(-1\pm\frac{1}{5}\right)$. They are real and negative, so the critical point is a stable node.

 For $y=\frac{V}{2}+ \frac{3\varepsilon^4}{25\lambda}$ the roots $\frac{\varepsilon^2}{2\lambda}\left(-1\pm\frac{7}{5}\right)$ are real and of different signs; the critical point is a saddle. The graph \ref{g1} of the solution on the phase plane starts from this point (the right one) as a separatrix.

\section{Nonlinear superposition}

Let us take the sum of the two waves of the type (\ref{TWS}) as the initial datum
\begin{equation}\label{sum}
 u(x.0)= TWS_1(x-V_1t+s_1)+TWS_2(x-V_2t+s_2)|_{t=0}.
\end{equation}

The limits values (at $\xi=\pm \infty$) adds,
\[H=H_1+H_2=\frac{V_1+V_2}{2}+\frac{6\varepsilon^4}{25\lambda},\;
h=\frac{V_1+V_2}{2}-\frac{6\varepsilon^4}{25\lambda}.\]

\begin{figure}[h]
\begin{minipage}{13.2pc}
\includegraphics[width=13.2pc]{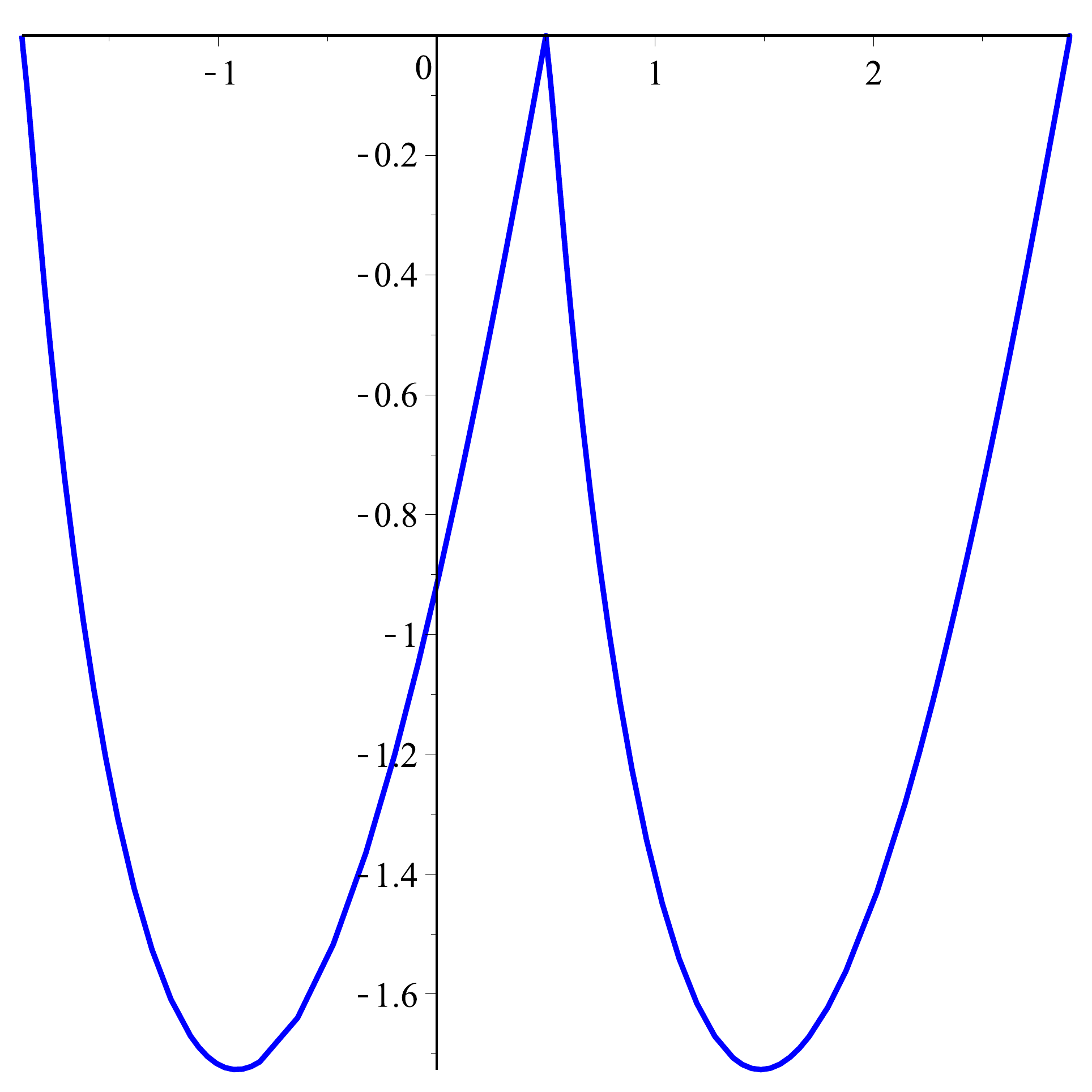}
\end{minipage}\hspace{2pc}%
\begin{minipage}{13.2pc}
\includegraphics[width=13.2pc]{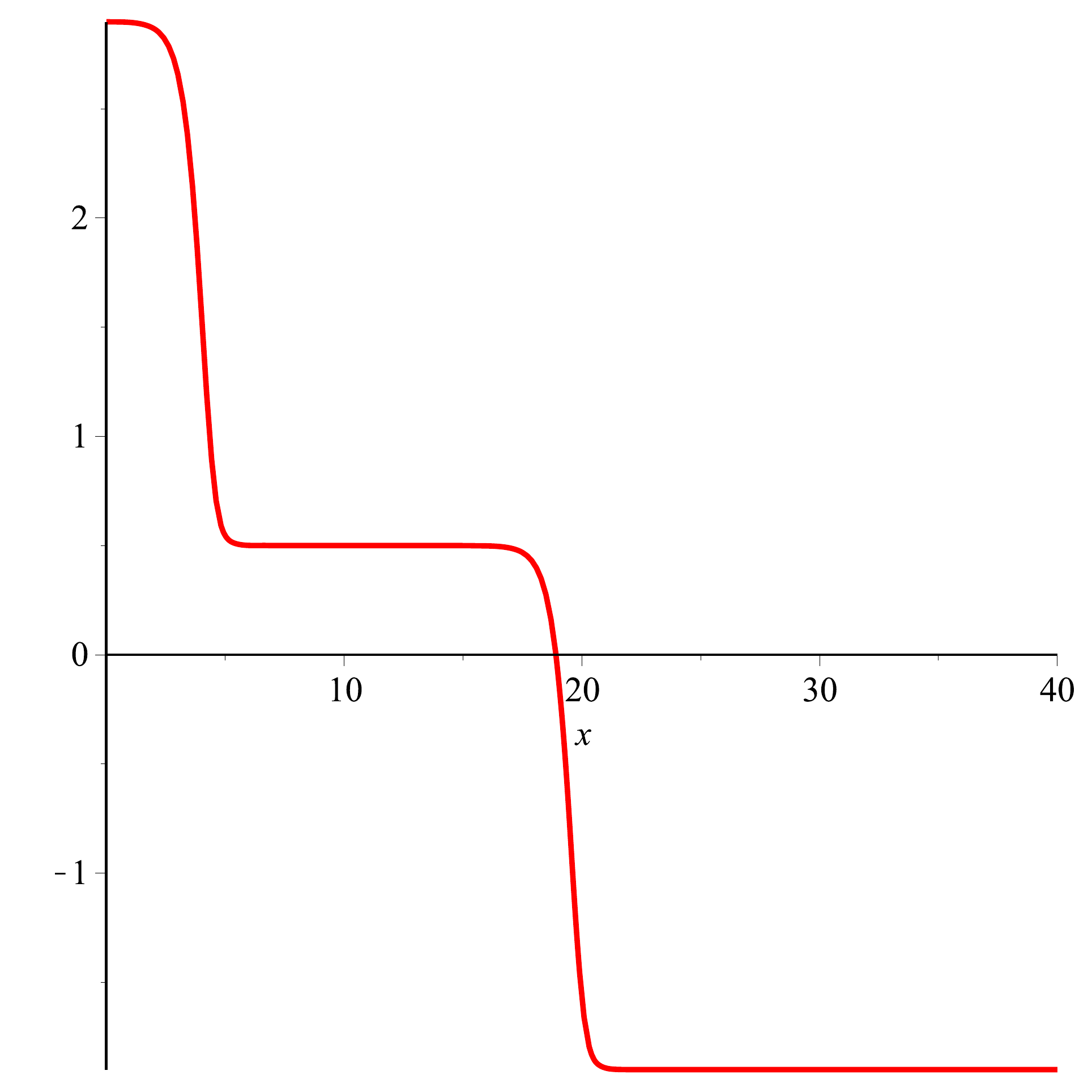}
\end{minipage}
\caption{\label{g2}Sum of shock waves\\  $s_1=4,\,V_1=5,\;s_2=20,\,V_2=-4,\;t=0$}
\end{figure}

 The initial value profile and the corresponding phase portrait at $t=0$ are given on figure \ref{g2}.

 The solution of  KdV-B with such an initial datum can not have a form (\ref{TWS}) since it has an unappropriate height ($12\varepsilon^4/25\lambda$,  instead of required $6\varepsilon^4/25\lambda$). However this solution at $t\rightarrow\infty$ becomes a (solitary) shock wave traveling with  the velocity $V=V_1+V_2$, \cite{key-4}.

\section{Transformation of the initial profile and reorganization of the phase portrait}

\begin{figure}[h]
\begin{minipage}{13.2pc}
\includegraphics[width=13.2pc]{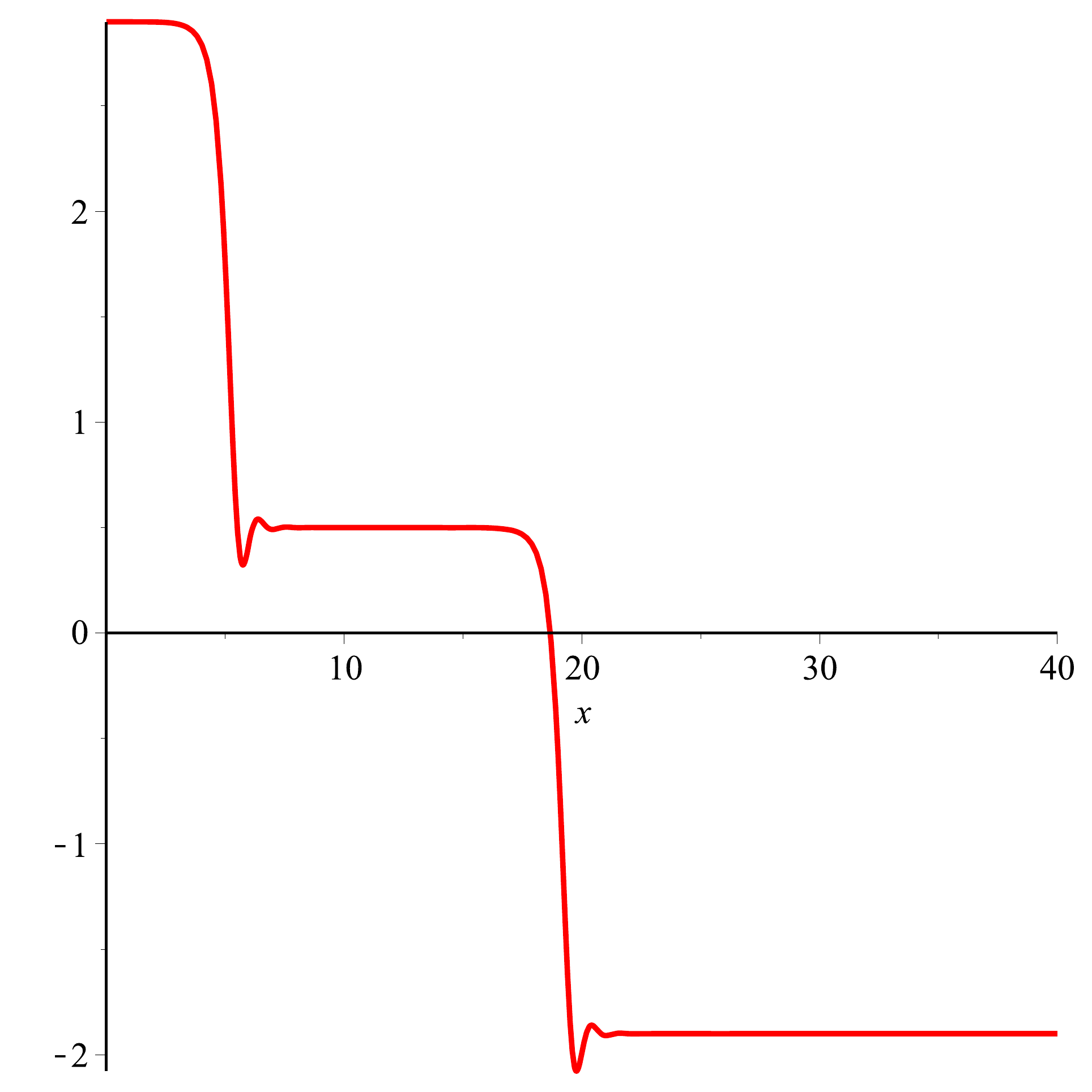}
\end{minipage}\hspace{2pc}%
\begin{minipage}{13.2pc}
\includegraphics[width=13.2pc]{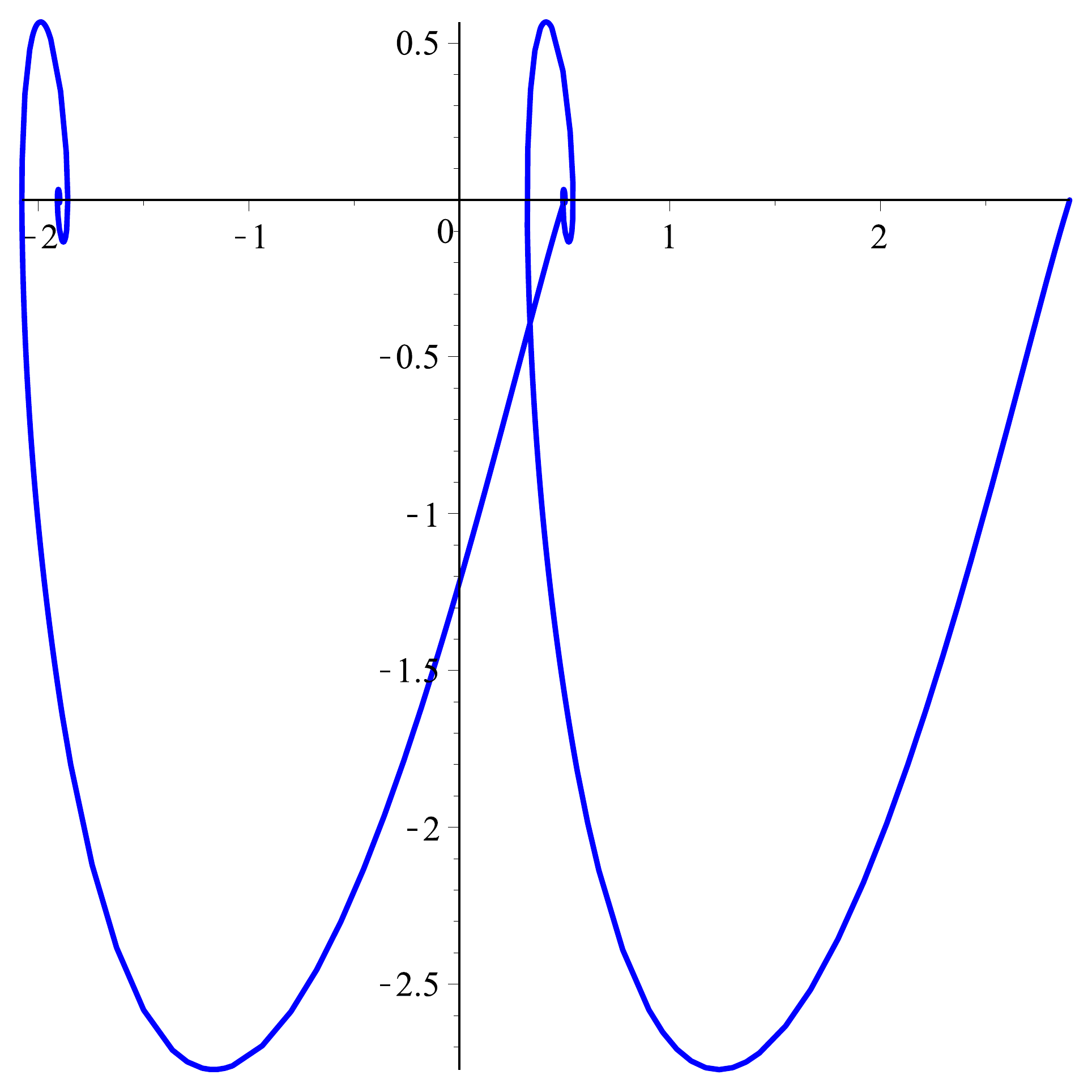}
\end{minipage}
\caption{\label{g3}Evolution of the sum of two shock waves,  $t=0.4$.\\
Profile and phase portrait}
\end{figure}

\begin{figure}[h]
\begin{minipage}{13.2pc}
\includegraphics[width=13.2pc]{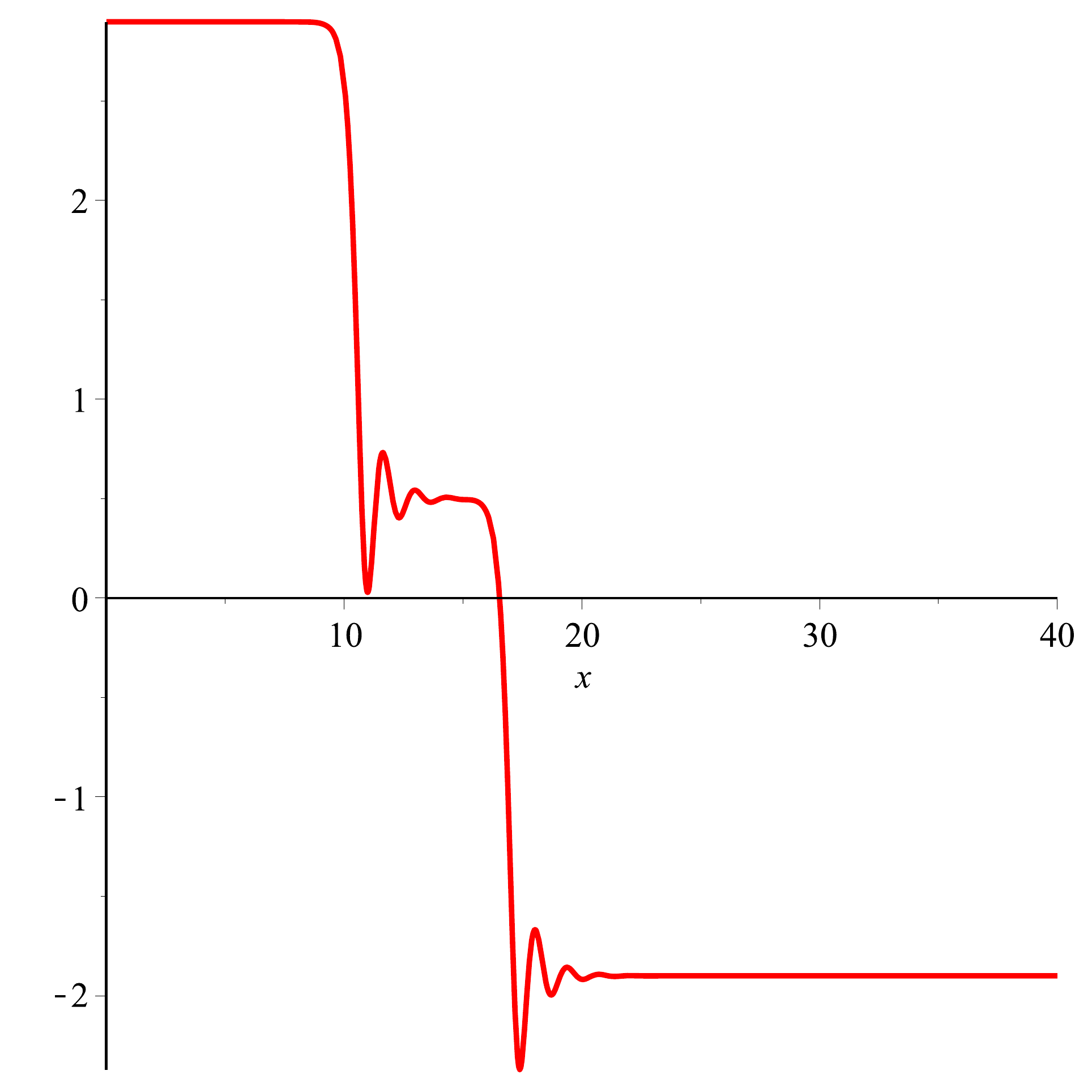}
\end{minipage}\hspace{2pc}%
\begin{minipage}{13.2pc}
\includegraphics[width=13.2pc]{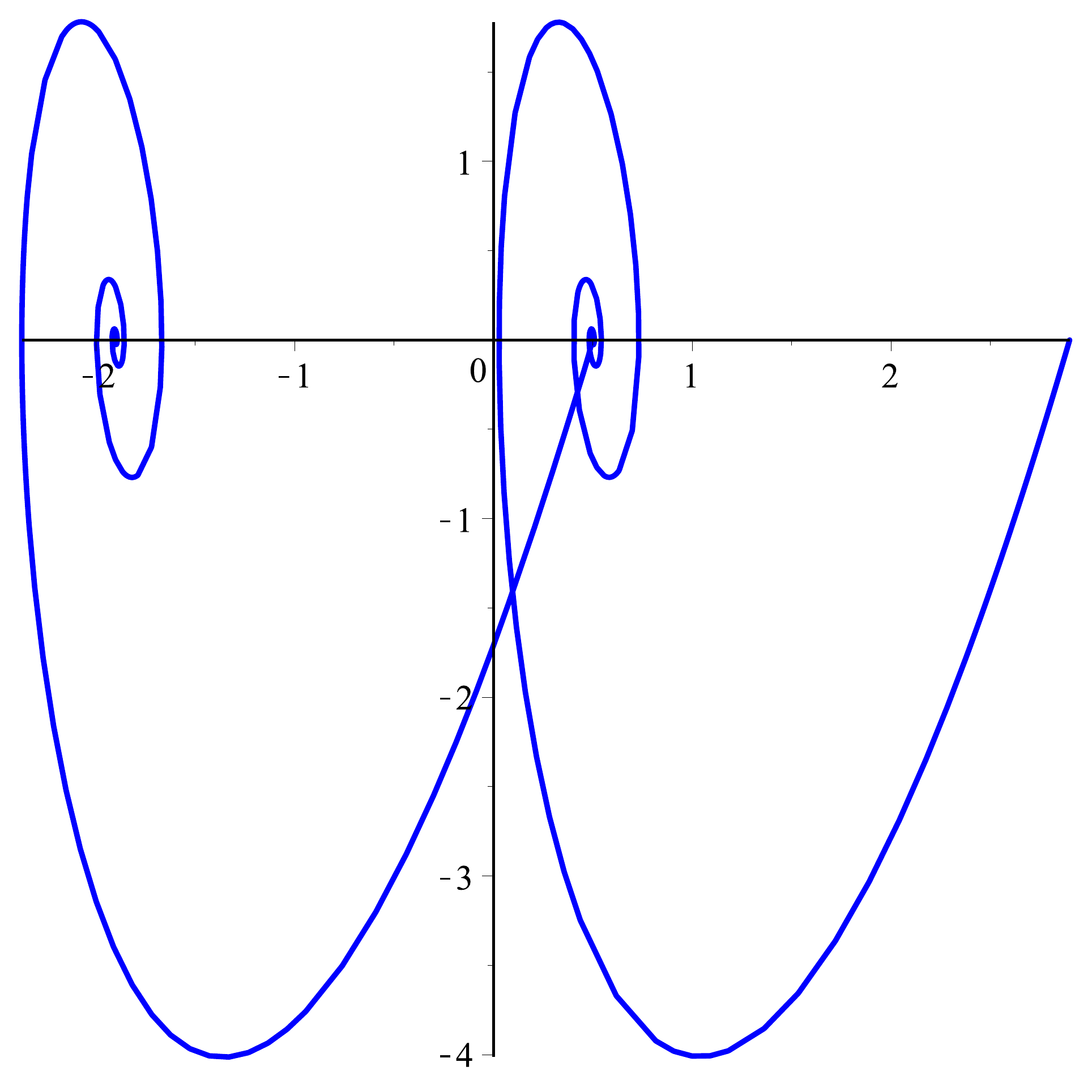}
\end{minipage}
\caption{\label{g4}Evolution of the sum of two shock waves,  $t=2$.\\
Profile and phase portrait}
\end{figure}

\begin{figure}[h]
\begin{minipage}{13.2pc}
\includegraphics[width=13.2pc]{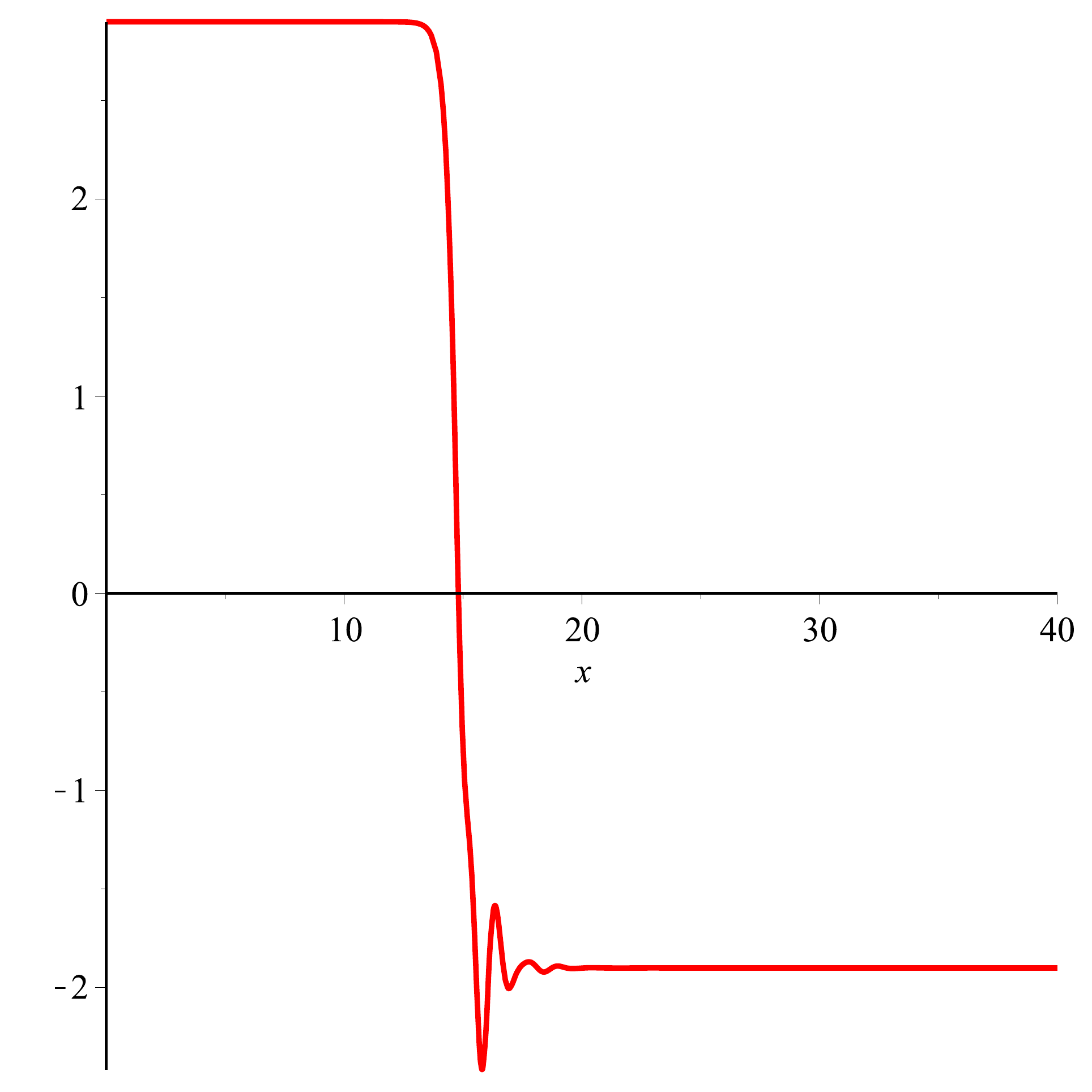}
\end{minipage}\hspace{2pc}%
\begin{minipage}{13.2pc}
\includegraphics[width=13.2pc]{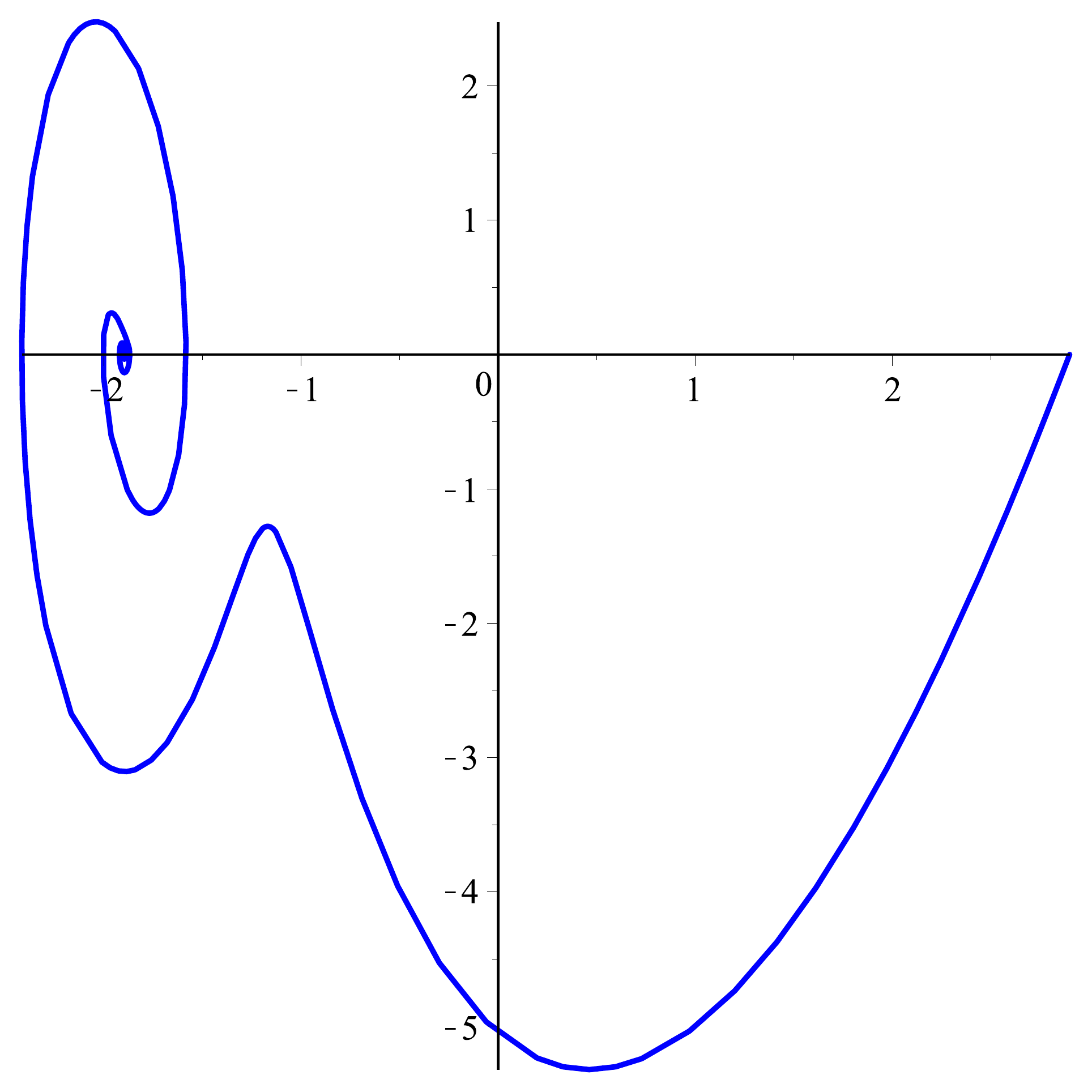}
\end{minipage}
\caption{\label{g5}Evolution of the sum of two shock waves,  $t=3.2$.\\
Profile and phase portrait}
\end{figure}

The process of evolution from the two-steps initial profile to a shock wave is illustrated by figures \ref{g2}--\ref{g5}; accompanied with the corresponding reorganization of the phase portrait. Initially these waves move towards each other (from $x=4$ to the right  with a velocity $V=5$ and from $x=20$ to the left with a velocity $V=4$). After the collision, beginning approximately at $t=3.3 $ a solitery shock wave is formed, moving to the right with a velocity   $V=1$.

\begin{figure}[h]
\begin{minipage}{13.2pc}
\includegraphics[width=13.2pc]{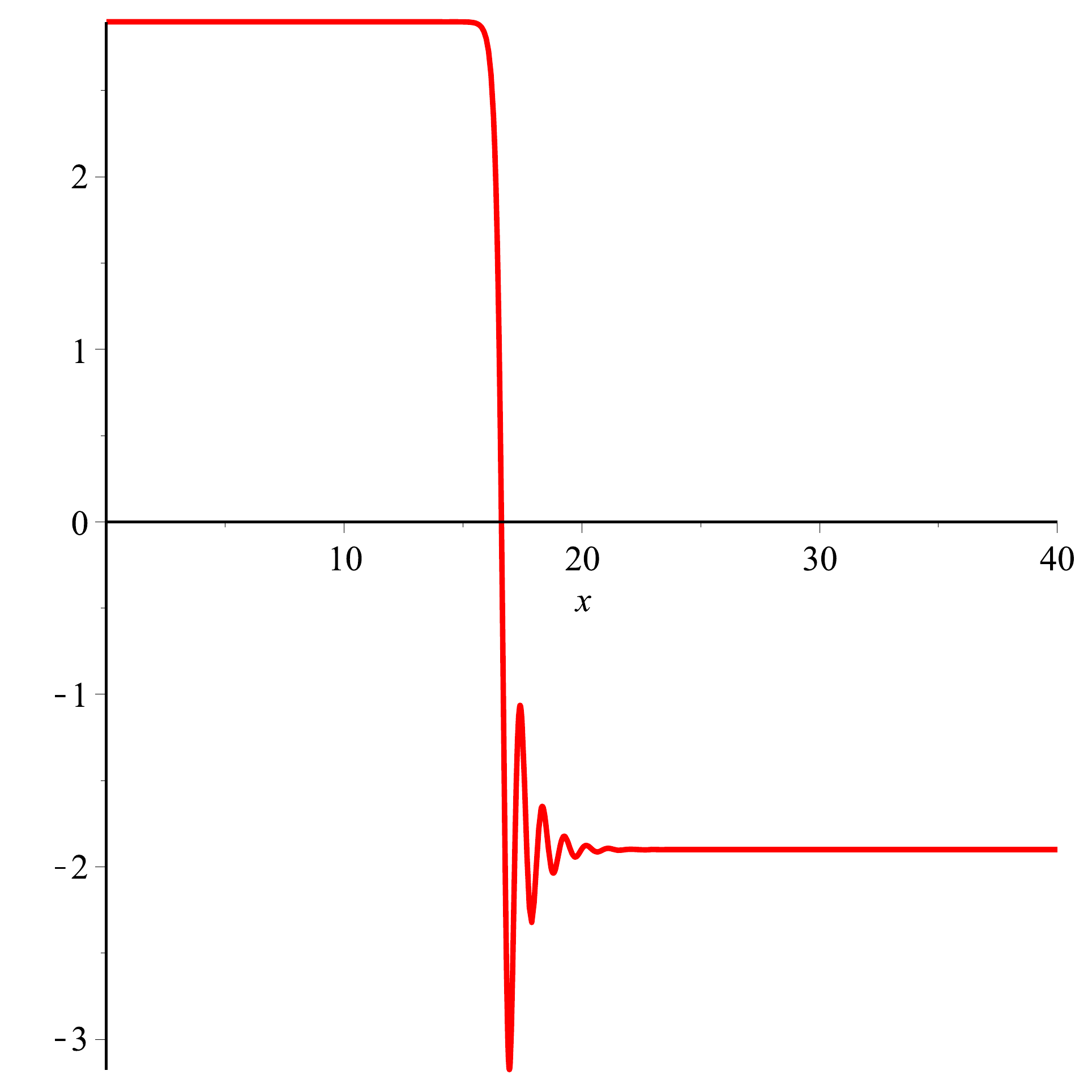}
\end{minipage}\hspace{2pc}%
\begin{minipage}{13.2pc}
\includegraphics[width=13.2pc]{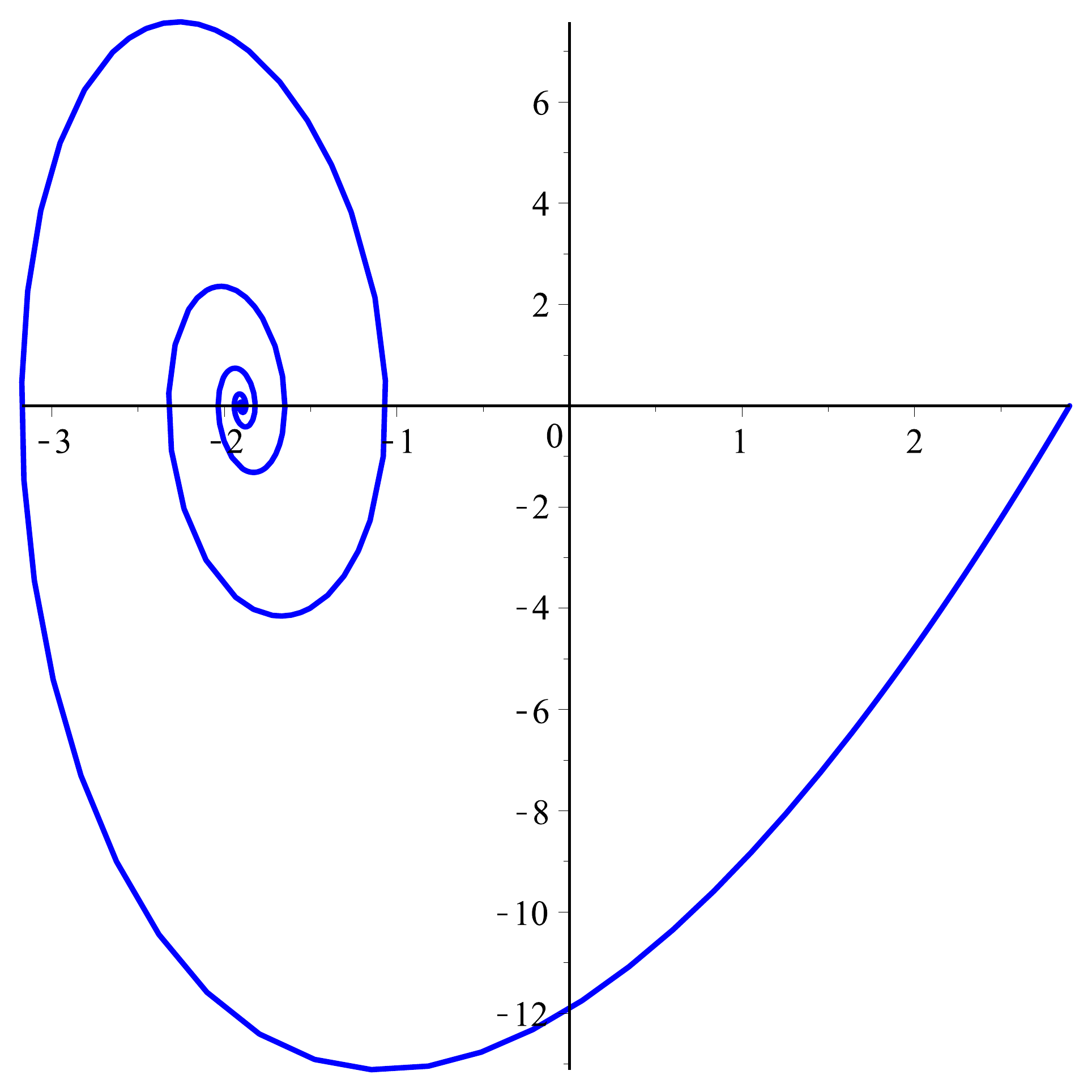}
\end{minipage}
\caption{\label{g6}Evolution of the sum of two shock waves,  $t=5$.\\
Profile and phase portrait}
\end{figure}

These graphs were obtained in the case $\lambda=0.1,\;\varepsilon=0.5$ and for the initial profile
\[\left\{\begin{array}{ccl}
    u(x,0) & = &(0.6\tanh(x-5t-4)^2-1.2\tanh(x-5t-4)+ \\
    & + & \left.0.6\tanh(x+4t-20)^2-1.2\tanh(x+4t-20)-0.7)\right|_{t=0}, \\
    u(0,t) & = & \left.u(x,0)\right|_{x=-\infty}=2.9.
  \end{array}\right.
\]

As it can be seen on the above graphs the lower limit level $y=h$ corresponds  to a stable focus. While approaching the collision moment the intermediate (approximate) fixed point also has a character of a stable focus.

Indeed, solving the characteristics equation \eqref{qe} for $y=\frac{V}{2}+ \frac{6\varepsilon^4}{25\lambda}$ we get the roots $\frac{\varepsilon^2}{2\lambda}\left(-1\pm\frac{\sqrt{73}}{5}\right)$.  They are real real and of different signs so the fixed point $y=\frac{V}{2}+ \frac{6\varepsilon^4}{25\lambda}$ is a saddle.

For $y=\frac{V}{2}- \frac{6\varepsilon^4}{25\lambda}$ the roots $\frac{\varepsilon^2}{2\lambda}\left(-1\pm\frac{\sqrt{23}}{5}i\right)$ are complex with a negative real part: the fixed point is a stable focus.

Adding three or more waves of the type \eqref{TWS} we obtain a similar picture. The solution at $t\rightarrow\infty$ becomes a solitary shock wave of the height $n\delta$ where $\delta=\frac{6\varepsilon^4}{25\lambda}$ is the height of the wave \eqref{TWS} and $n$ is a number of summands. The more summands the greater the frequency of oscillations preceding the forward front of the forming shock wave.

Thus, because of  dissipation,  after adding several arbitrary and localized shock waves  one can expect that the sum's asymptotic is also a shock wave which velocity and height are predictable.  The behavior of the solution in vicinity of fixed points or  at the time of the summands' colliding can be described as above.

In particular, for numerical simulation it is convenient to take summands of the type \eqref{TWS} with parameters $\lambda,\varepsilon$ that differ from those of the equation.

Note that constant $K$ may be considered as a shock wave  of a zero height and a velocity $2K$.
Hence adding a constant $K$ to a solution (\ref{TWS}) increases  its velocity by  $2K$.  The fixed points are shifted by $K$:
 $y =\frac{V}{2}\pm \frac{3\varepsilon^4}{25\lambda}+K$. This explains the  intermediate (approximate) fixed point behavior  in vicinity of the collision.

\section{Inference and numerical considerations}

It is impossible to write down an exact solution to the KdV-B so it is especially important to study its invariant solutions (with respect to symmetries). Surely exact invariant solutions are very particular and it is not clear whether they arise in the course of the evolution of an arbitrary initial problem. Yet the practice and numerical modeling demonstrate that many of them play an important role, being  a sort of an attractor and/or a separatrix: the behavior of most solutions at $t\rightarrow\infty$ coincides as a rule with that of an invariant solution.  End numeric simulation gives us an invaluable possibility to understand the forming of solutions in detail.

The graphs in this paper were obtained by  numerical methods using  the Maple PDETools package. It is worth to note that multi-oscillating is an intrinsic  property for the KdV-B equation and a spatial derivative may change abruptly. In this situation the standard methods used with default parameters  may easily loose stability, leading to a general loss of precision. We dealt with this problem mostly adapting parameters   \emph{spacestep} and/or \emph{timestep} of the PDETools package methods.


\begin{thebibliography}{1}

\bibitem[1] {key-1} \textit{A.P. Chugainova, V.A. Shargatov,} Stability of the breaks structure described by the generalized Kortweg-de Vries-Burgers equation, Computational Math and Math Phys. 56:2 (2016), 259–274

\bibitem[2]{key-2} \textit{B. Dubrovin, M. Elaeva.} On critical behavior in nonlinear evolutionary PDEs with small viscosity. ArXiv: 1301.7216v1math-ph., 30.01.2013, 16 p3.

\bibitem[3]{key-3}    \textit{B. Dubrovin.} On Hamiltonian Perturbations of Hyperbolic Systems of Conservation Laws, II: Universality of Critical Behaviour, Comm. Math. Phys.,  267 (2006), pp. 117–-139.

\bibitem[4]{key-4} \textit{N.M. Ryskin, D.I. Trubetskov,} Nonlinear waves - Moscow, Fizmatlit, 2001. - 314 p. (in Russian).

\bibitem[5]{key-5} \textit{A. Samokhin.} On  Burgers Equation with a Periodic Boundary Conditions on an Interval, J. Theor. Math. Phys., 168:4, (2016) (to appear).

\bibitem[6]{key-6} \textit{ O.V. Rudenko.} Nonlinear sawtooth-shaped waves. UFN, \textbf{9} (1995), pp. 1011–-1035 (in Russian).
\end{thebibliography}
\end{document}